\documentclass[letterpaper,twocolumn,fleqn]{article} 

\usepackage{ist}

\usepackage[ruled,vlined]{algorithm2e}
\usepackage{array}
\usepackage{multirow}
\usepackage{tabularx}
\usepackage{amsfonts}
\usepackage{algcompatible}
\usepackage{amsmath}

\title{A Metal Artifact Reduction Scheme For Accurate Iterative Dual-Energy CT Algorithms}

\author{ Tao Ge\textsuperscript{1}, Maria Medrano\textsuperscript{1}, Rui Liao\textsuperscript{1}, Jeffrey F. Williamson\textsuperscript{2}, David G. Politte\textsuperscript{3}, Bruce R. Whiting\textsuperscript{4}, Joseph A. O’Sullivan\textsuperscript{1} \\
\textsuperscript{1}Department of Electrical \& Systems Engineering, Washington University in St. Louis, St. Louis, MO\\
\textsuperscript{2}Department of Radiation Oncology, Washington University in St. Louis, St. Louis, MO\\
\textsuperscript{3}Mallinckrodt Institute of Radiology, Washington University in St. Louis, St. Louis, MO\\  \textsuperscript{4}Department of Radiology, University of Pittsburgh, Pittsburgh, PA}

\date{} 

\begin{document} 

\maketitle 

\thispagestyle{empty} 


\begin{abstract}
CT images have been used to generate radiation therapy treatment plans for more than two decades. Dual-energy CT (DECT) has shown high accuracy in estimating electronic density or proton stopping-power maps used in treatment planning. However, the presence of metal implants introduces severe streaking artifacts in the reconstructed images, affecting the diagnostic accuracy and treatment performance. In order to reduce the metal artifacts in DECT, we introduce a metal-artifact reduction scheme for iterative DECT algorithms. An estimate is substituted for the corrupt data in each iteration. We utilize normalized metal-artifact reduction (NMAR) composed with image-domain decomposition to initialize the algorithm and speed up the convergence. A fully 3D joint statistical DECT algorithm, dual-energy alternating minimization (DEAM), with the proposed scheme is tested on experimental and clinical helical data acquired on a Philips Brilliance Big Bore scanner. We compared DEAM with the proposed method to the original DEAM and vendor reconstructions with and without metal-artifact reduction for orthopedic implants (O-MAR). The visualization and quantitative analysis show that DEAM with the proposed method has the best performance in reducing streaking artifacts caused by metallic objects.
\end{abstract}

\section{Introduction}
\label{sec:intro}

Compared to single-energy CT, Dual-Energy CT (DECT) is a computed tomography technique that generates more informative images from data acquired at different x-ray photon energy spectra. DECT images have been widely used for diagnosis and therapy guides for decades. For instance, \cite{Maria20} show that the iterative DECT algorithm achieved sub-percentage error in estimating electron density maps and proton stopping-power maps for the proton therapy treatment plan.

However, due to the fast development of modern medicine, medical implants are widespread in patients, especially in older people. There are different types of implants, including dental implants, dental crowns, and spine implants. Most of the implants contain metal, usually titanium or stainless steel. The presence of metal implants introduces severe streaking artifacts in the CT reconstruction since metallic objects significantly attenuate and scatter the X-ray beam passing through it, reducing the probability of photons being received by the target detector. 

The mean energy of the polychromatic beam increases after x-ray beam penetrates the high-attenuation object, producing the beam-hardening effect. The denser object is also more likely to change the photon path, resulting in additional scatter. Since the number of transmitted photons obeys the Poisson distribution, the metal object would significantly decrease the probability that the detector receives the photon. Moreover, the projection of a high-contrast boundary to a single detector introduces the nonlinear partial volume effect (NLPV) \cite{7565564}. In dual-energy CT, the misalignment of high-attenuation objects could also lead to severe streaking artifacts.

Metal-artifact reduction (MAR) techniques have been developed for decades. Current mainstream MAR methods can be classified into three types, traditional pre-reconstruction correction, iterative reconstructions, and deep learning. Pre-reconstruction methods could be further divided into physics-based correction (noise filtering, scatter or Beam-Hardening Correction) and sinogram completion that attempts to interpolate the data corrupted by metallic objects. One of the well-known completion techniques is called NMAR, where the sinogram is interpolated after the normalization \cite{Meyer2010}. The metal artifact could also be reduced by omitting the corrupted data \cite{538943} or introducing a more realistic forward model \cite{959297} during the reconstruction. The basic idea is to down-weight the corrupted data in the update procedure. Moreover, the introduction of the polychromatic transmission model could eliminate the beam-hardening artifact. Numerous deep-learning-based methods for MAR have been published in recent years. The dual-domain networks are a type of popular end-to-end trainable network designed to reduce the metal artifact in both the sinogram and the image domain \cite{Lin_2019_CVPR,8815915,9201079}.  

Previous literature regards the iterative polyenergetic CT algorithm as a solution to reduce the streaking artifact caused by the presence of metal in the field of view. However, the topic of reducing metal artifacts in the iterative polyenergetic CT algorithms has not been fully discussed. In this work, we propose a MAR framework for iterative DECT algorithms. The corrupted-data omission method is applied to an iterative DECT algorithm with the polyenergetic statistical model, and the sinogram completion method is used for initialization. The proposed framework is tested on both experimental and clinical data, and the performance of the proposed method is quantitatively assessed. In the experimental test, the image reconstructed by the proposed method is compared to the vendor reconstruction with MAR.

\section{Methods}

\subsection{DECT with MAR}
The first MAR method treats the corrupt data as missing. It follows the idea of reconstructing the incomplete transmission data \cite{Snyder2006ImageRF}. For any iterative algorithm with a general data fidelity term 
\begin{equation}
\min \sum_{j=L,H} \sum_y D\Big(d_j\big(y\big), g_j\big(y:(c_1,c_2)\big)\Big),\\
\label{eq:data}
\end{equation}
one could always substitute the measured data $d_j\big(y\big)$ as
\begin{align}
\min& \sum_{j=L,H} \sum_y D\Big(\tilde{d}_j\big(y\big), g_j\big(y:(c_1,c_2)\big)\Big)\\
\text{subject to}&  \quad \tilde{d}_j(y)=d_j\big(y)\big)\quad  T(y)<t
\end{align}
where $D:R^N_{xyz}\to R$ is the point-wise data-fit function, $y$ denotes the measurement index, $d_L$ and $d_H$ denote the low- and high-energy transmission data, respectively, $T(y)=\sum_x h(x,y) M(x)$ is the metal trace sinogram generated by forward-projecting the metal-only image $M(x)$, and $t$ is a threshold scalar, $\tilde{d}_j(y)$ is the nonnegative substitution of the measured data whose value is arbitrary when $T(y)\geq t$, and $c_1$, $c_2$ denote two material decomposition components. Note that in practice, the system operators may be different because of the scanner settings or the object motion during scanning. Then, different metal traces $T_j(y)$ are required to match the low- and high-energy data, respectively. Here we assume the system operators are the same to simply the problem.

\subsection{MAR Initialization}

\begin{figure}[!t]
  \centering
  \includegraphics[width=0.47\textwidth]{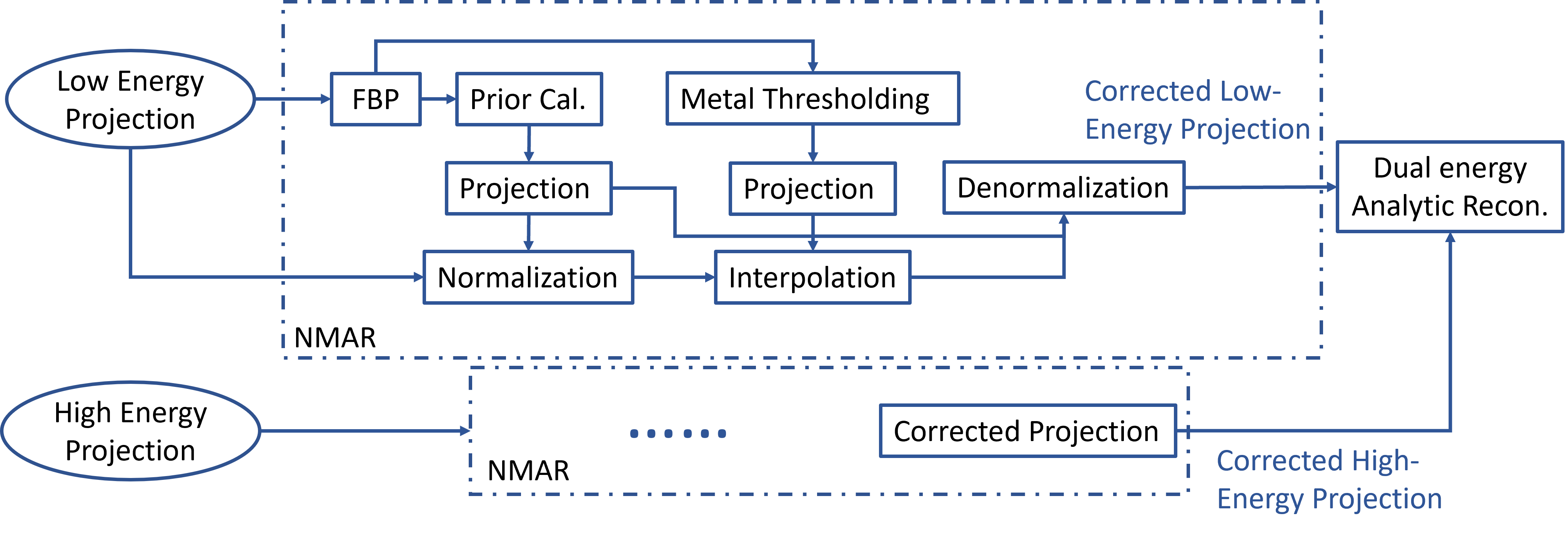}
  \caption{Flowchart of MAR Initialization.}
  \label{fig:flowchart}
\end{figure}

Down-weighting or ignoring the measured data in the metal trace leads to smaller updating steps in the metal region, resulting in the low convergence rate of iterative DECT algorithms. The MAR initialization method provides MAR-DECT with the metal-artifact-corrected initial condition, helping MAR-DECT to converge faster. The flowchart of the MAR Initialization method is shown in Figure \ref{fig:flowchart}. In this work, the initial image is evaluated using the analytic DECT reconstruction method cooperating with the normalized metal-artifact reduction (NMAR) method \cite{Meyer2010} from measured sinograms acquired with low- and high-energy spectra. 

In NMAR, before the interpolation of the metal trace, the measured projection data is normalized to the forward-projection of the prior image. The interpolated sinogram is denormalized to generate the corrected sinogram. With a proper prior image, this method is capable of preserving the edge information in the projection sinogram while interpolating the corrupted phantom beams. Note that the corrected sinograms are exclusively used in the initialization process of this MAR scheme.

Image-domain decomposition is an analytic DECT reconstruction method that evaluates basis-component images from two filtered back-projection (FBP) images of low- and high-energy projections. One could easily get the basis-component images via the linear combination of two FBP images \cite{shuangyue2018}, 

 \begin{equation}
\begin{bmatrix}
c_1(x)\\
c_2(x)
\end{bmatrix}
= 
\begin{bmatrix}
\tilde{\mu}_{1,L}(x) & \tilde{\mu}_{2,L}(x)\\
\tilde{\mu}_{1,H}(x) & \tilde{\mu}_{2,H}(x)
\end{bmatrix}^{-1}
\begin{bmatrix}
\mu_L(x)\\
\mu_H(x)
\end{bmatrix}, 
\end{equation}
where $\tilde{\mu}_{i,j}(x)$ denotes the modified weights obtained from calibration images.

Besides the image-domain decomposition method, the sinogram domain decomposition or iterative filtered back-projection could also be utilized in this step. 

\begin{figure}[!t]
  \centering
  \includegraphics[width=0.47\textwidth]{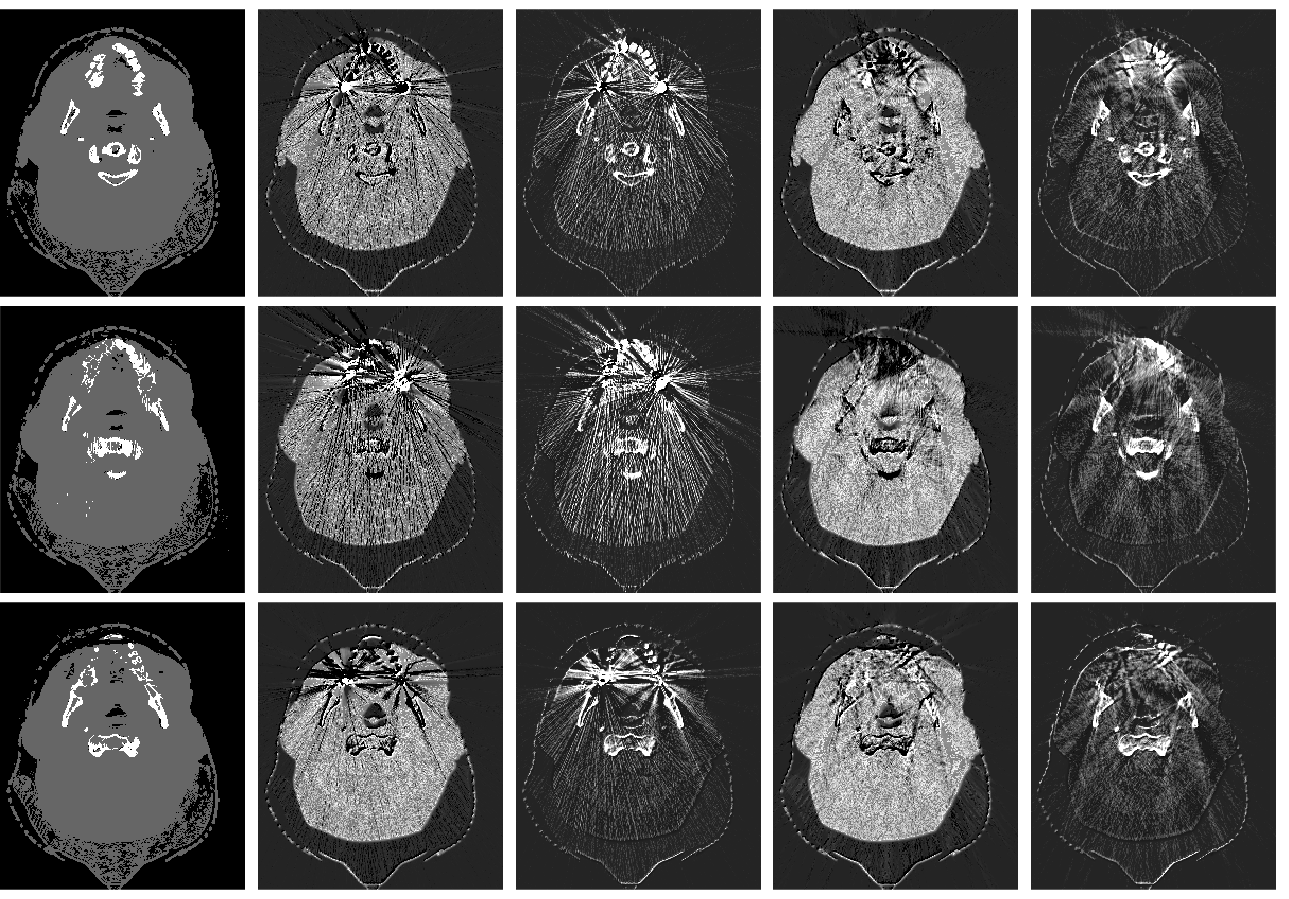}
  \caption{An example of real-patient MAR initialization. Column 1: Prior image by thresholding and filtering. Columns 2,3: $c_1$ and $c_2$ of FBP-based image-domain decomposition result without MAR, respectively. Column 4,5: $c_1$ and $c_2$ of FBP-based image-domain decomposition result with NMAR, respectively. Rows 1-3: different slices (z-direction index).}
  \label{fig:NMAR_patient}
\end{figure}

An example of clinical MAR initialization images is shown in Figure \ref{fig:NMAR_patient}. The prior image is generated by thresholding the single-energy FBP image.

\subsection{MAR-DEAM}
In this section, we take the Dual-Energy Alternating Minimization (DEAM) algorithm as an example to illustrate the implementation details of the MAR framework. 

DEAM is a joint statistical iterative DECT algorithm that minimizes the objective function given by the sum of I-divergence \cite{o07}, 
\begin{equation}
I(d||g)=\sum_j \sum_y \Big(d_j(y)\ln\frac{d_j(y)}{g_j(y:c)}-d_j(y)+g_j(y:c)\Big),
\end{equation}
and a penalty term, 
\begin{equation}
R(c)=\lambda \sum_{i=1}^2\sum_x\sum_{\tilde{x}\in N_x} w(x,\tilde{x}) \varphi\left(c_i(x)-c_i(\tilde{x})\right),
\end{equation}
\begin{equation}
\varphi(t) = \delta^2\left(\left|\frac{t}{\delta}\right|+\log\left(1-\left|\frac{t}{\delta}\right|\right)\right),
\end{equation}
where $x$, $y$ denote the indices of the discretized image space and measurement space, respectively. $N_x$ denotes the set of the neighboring voxels of the image index $x$, $w(x,\tilde{x})$ is the voxel weight calculated as the inverse physical distance between voxel $x$ and $\tilde{x}$, $\lambda$ and $\delta$ are two hyper-parameters that control the weight and sparsity of the regularization term. $i$ denotes image component index, $j$ denotes measured data index, $d$ denotes measured data, and $g(y:c)$ denotes the estimation of measured data based on image components $c_i$, which is the forward model, written as
\begin{equation}
g_j(y:c)=\sum_E I_{0,j}(y,E)\exp\left(-\sum_x h(x,y)\sum_{i=1}^2 \mu_i(E)c_i(x)\right),
\end{equation}
where $\mu_i(E)$ denotes the attenuation coeffcient of the $i^{th}$ material at energy $E$, $I_{0,j}$ denotes the photon counts of the $j^{th}$ peak energy in the absence of an object, which contains information of the spectrum and the bowtie filter, and $h(\cdot,\cdot)$ denotes the system operator that represents the helical fan-beam CT system.

To implement the first proposed MAR method, $d$ is substituted with $\tilde{d}$, where $\tilde{d}(y)=d(y)\quad \forall \{y|T(y)<t\}$, and $\tilde{d}(y)$ has arbitrary value for $\{y|T(y)\geq t\}$. From O’Sullivan and Benac \cite{o07}, the I-divergence between $d$ and $g$ is equivalent to minimizing the energy-summed I-divergence between single-energy sinograms $p$ and $q$ over $p$, subject to the linear family. Specifically, 

\begin{multline}
\sum_j I(\tilde{d}_j||g_j(c))= \min_{p_j\in \ell(\tilde{d})} \sum_j\sum_y \sum_E\\
 \Big(p_j(y,E)\ln\frac{p_j(y,E)}{q_j(y,E)}-p_j(y,E)+q_j(y,E)\Big),
\end{multline}
where $\ell(\tilde{d})$ is the constraint of the linear family
\begin{equation}
\ell(\tilde{d}_j)=\Big\{p_j(y,E)\geq 0\Big| \sum_E p_j(y,E)=\tilde{d}_j(y)\Big\}.
\end{equation}

Then, substituting $\tilde{t}$ with $t$ gives
\begin{multline}
I(\tilde{d}||g(c))=\min_{p(y)\in \ell(d(y))}  \sum_E\\
 \Big(\sum_{\{y|T(y)<t\}}\big(p_j(y,E)\ln\frac{p_j(y,E)}{q_j(y,E)}-p_j(y,E)+q_j(y,E)\big)\\
+\sum_{\{y|T(y)\geq t\}} \big(p_j(y,E)\ln\frac{p_j(y,E)}{q_j(y,E)}-p_j(y,E)+q_j(y,E)\big)\Big),
\end{multline}
which yields $p_j(y,E)=q_j(y,E)$  $\forall E, j$ and $\{y|T(y)\geq t\}$. In other words, in MAR-DEAM, $p_j(y,E)$ will be substituted with $q_j(y,E)$ in each iteration, if the index $y$ satisfies $T(y)\geq t$.

\section{Results}
\subsection{Experimental Results}

\begin{figure}[!t]
  \centering
  \includegraphics[width=0.48\textwidth]{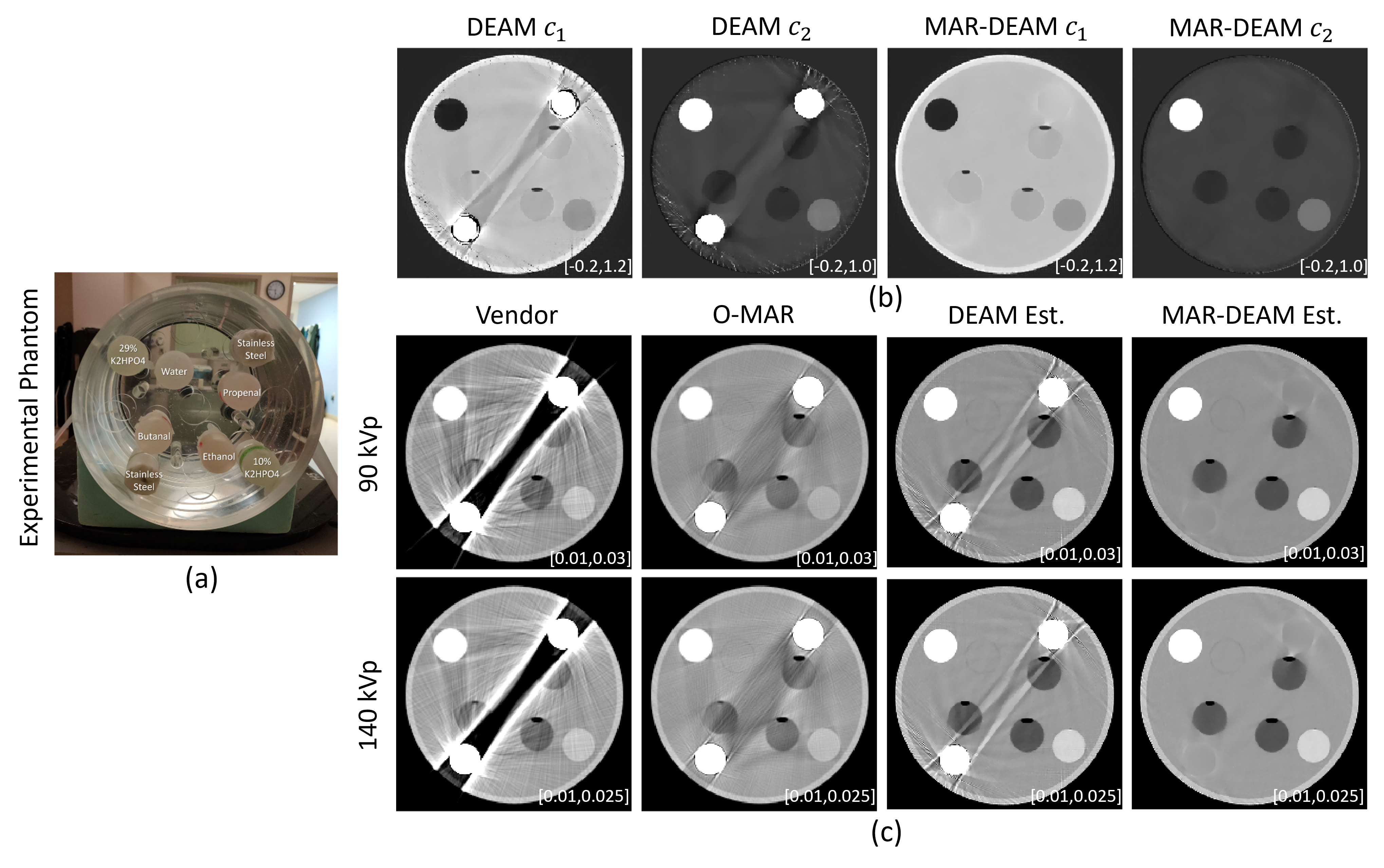}
  \caption{(a) The head water phantom with 12 inserts used in the experimental analysis, (b) Basis-component images, (c) 90 and 140 kVp reconstructed images by vendor and DEAM.}
  \label{fig:phantom}
\end{figure}

In the experimental test, we used a cylindrical water phantom with six different inserts to simulate the property of tissues in the human body (shown in Figure \ref{fig:phantom}(a)). Two stainless-steel rods are placed on the diagonal, so representative soft-tissue samples (butanol and propanol) are directly affected. The inserts have a diameter of 32 mm, and the stainless-steel rod has a diameter of 24 mm. The experimental data was acquired on a Philips Brilliance Big Bore scanner.
Figure \ref{fig:phantom}(b) shows the basis-component images reconstructed by DEAM and MAR-DEAM, respectively, and \ref{fig:phantom}(c) shows the 90 and 140 kVp reconstructed images by the vendor without metal-artifact reduction for orthopedic implants (O-MAR), the vendor with O-MAR, DEAM, and MAR-DEAM, respectively. The DEAM and MAR-DEAM images in \ref{fig:phantom}(c) are estimated from the basis components results by spectrum-weighted averaging the corresponding set of virtual mono-energetic images. O-MAR is an iterative MAR algorithm where the correction image is recursively subtracted from the original image that is evaluated by backprojecting the masked difference sinogram \cite{omar}.

From the $c_{1}-c_{2}$ images, we can see that MAR-DEAM reduces most of the streaking artifacts in the DEAM images. In the 90 and 140 kvp results, the vendor reconstruction without O-MAR performs the worst. The O-MAR images have fewer streaking artifacts than the vendor reconstruction without O-MAR and the original DEAM, but some artifacts are still present between the high-contrast inserts. In the DEAM image, noise is concentrated near the metal rods and is probably caused by photon starvation. 

By comparing the MAR-DEAM image to the others, it can be seen that the proposed method almost eliminates the metal artifact in the reconstructed image. However, note that since the measured data with the metal information is ignored, the proposed method is not able to reconstruct the metal region at all. This issue could be addressed by separately reconstructing the metal region.

\begin{figure}[!t]
  \centering
  \includegraphics[width=0.47\textwidth]{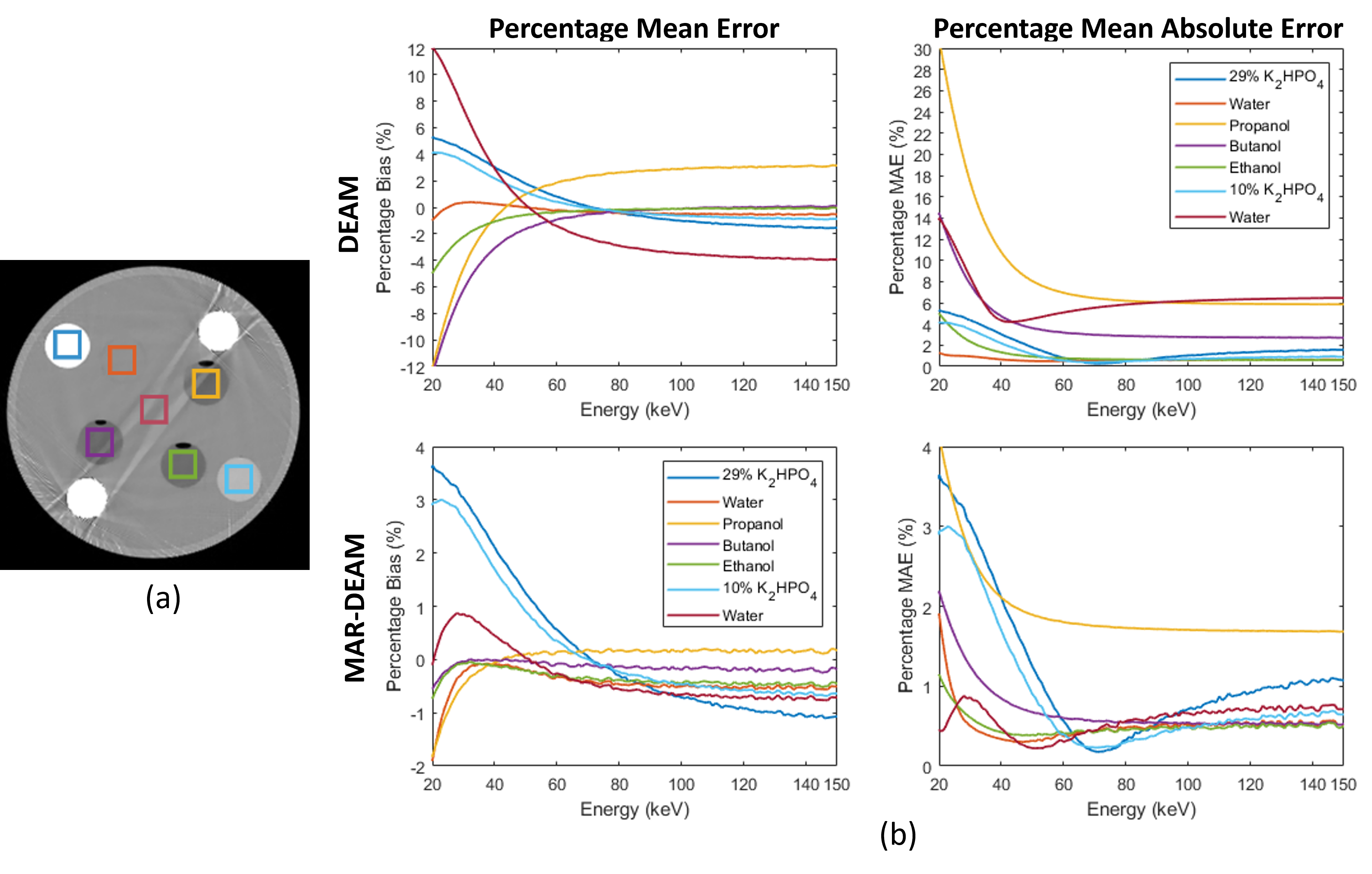}
  \caption{(a) ROIs are indicated by squares. (b) Percentage bias and mean absolute error (MAE) of virtual mono-energetic images from DEAM results. }
  \label{fig:simu}
\end{figure}

\begin{figure*}[ht!]
  \centering
  \includegraphics[width=0.95\textwidth]{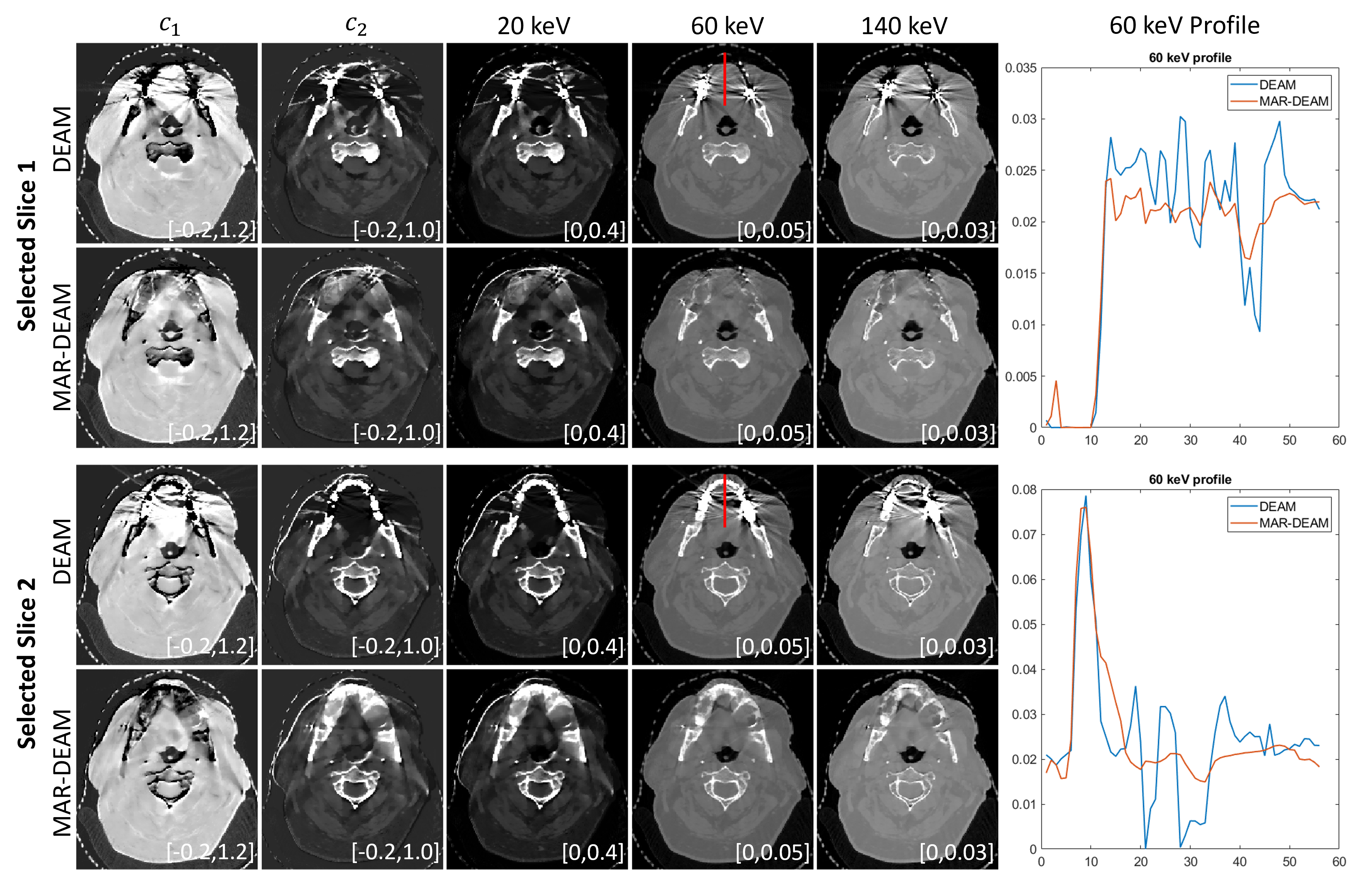}
  \caption{The MAR-DEAM results compared to DEAM. Column (1): first basis-component image $c_1$ (polystyrene) with the display window [-0,1.2]. Column (2): second basis-component image $c_2$ (CaCl$_2$). Column (3): virtual mono-energetic image at 20 keV generated from $c_1$ and $c_2$. Column (4): virtual mono-energetic image at 60 keV generated from $c_1$ and $c_2$. Column (5): virtual mono-energetic image at 140 keV generated from $c_1$ and $c_2$. Column (6): profiles of mono-energetic image at 60 keV. Row (1) and (2): DEAM result and MAR-DEAM result for the first selected slice, respectively. Rows (3) and (4): DEAM result and MAR-DEAM result for the second selected slice, respectively. }
  \label{fig:patient_image}
\end{figure*}
Figure \ref{fig:simu} shows the quantitative experimental result. Seven cubes of regions of interest are indicated in \ref{fig:simu}(a), and the image on the lower left shows the plot of linear attenuation coefficients of ICRU values of selected materials vs. energies. ICRU values will be used as ``truth" in this test. 
The first row in \ref{fig:simu}(b) shows the plot of percentage bias and percentage MAE of selected ROIs for the original DEAM, respectively, and the second row shows the plot of percentage bias and percentage MAE of selected ROIs for MAR-DEAM. From the bias plot for DEAM, although the biases in most of the ROIs are within +/-1\% from 60 keV to 150 keV, the biases of propanol and water at the center are relatively high. Since these two regions are between two stainless-steel rods, the metal artifact significantly affects the value. 

From the bias plot for MAR-DEAM, the percentage mean error of all ROIs is within 1\% from 60 keV to 150 keV except for the 29\% K$_2$HPO$_4$. In the original DEAM, propanol and water at the center are the two regions most affected by the metal artifact. Now the biases of these two regions are significantly reduced. 
This trend could also be seen in the percentage MAE plots. Comparison of the two plots in the right column shows that MAEs of all the materials are reduced by the proposed method, especially for propanol, butanol, and water at the center. Moreover, MAE of all the ROIs for MAR-DEAM is within 1\% for energies greater than 60 keV, except for propanol and 29\% K$_2$HPO$_4$.

\subsection{Clinical Results}

A representative head and neck patient was scanned at a 16$\times$0.75 mm collimation using a Philips Big Bore scanner. The selected data trunk has 816 columns and 10560 views, and the reconstructed image dimension is 610$\times$610$\times$85 mm.

Figure \ref{fig:patient_image} shows the reconstructed images from the patient data. From the left column to the right are $c_1$, $c_2$, 20 keV virtual monoenergetic images, 60 keV virtual monoenergetic images, and 140 keV virtual monoenergetic images. The plot on the right shows the vertical profile of the 60 keV DEAM and MAR-DEAM images. The profile region is indicated by the red line on the 60 keV image of DEAM. Two slices are selected in the reconstructed volume.

It can be seen that DEAM with the proposed method successfully reduces the streaking artifact caused by dental crowns. However, the proposed algorithm could not correctly reconstruct the metal region due to the ignored data. Ignoring the metal data does not just lead to the missing of the dental crowns, but it also produces an artifact in the missing region, which leads to an overestimation at 20 keV and underestimation at 140 keV, compared to the attenuation of the soft tissue. The value of the metal region is estimated from the initial image and will be gradually updated by the penalty term in the iterative algorithm.

\begin{figure}
  \centering
  \includegraphics[width=0.47\textwidth]{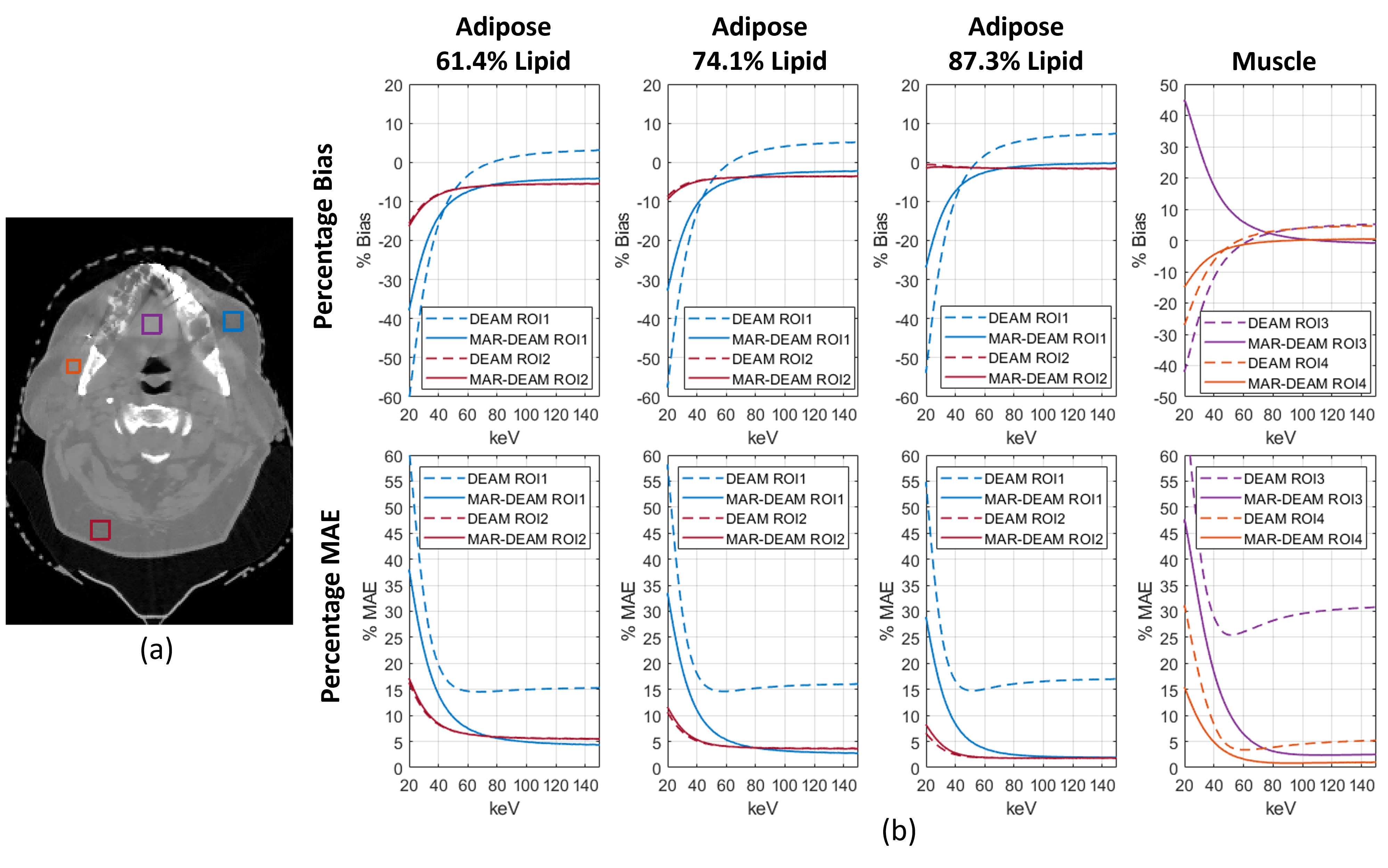}
  \caption{Quantitative results of DEAM and MAR-DEAM. (a): Three cubes are selected as the ROI for the quantitative analysis.The blue and yellow ROIs correspond to adipose tissue, and the material in the red ROI is recognized as muscle tissue (tongue). (b) Plots of percentage biases and MAEs of selected regions versus energy.}
  \label{fig:patient_plot}
\end{figure}

In figure \ref{fig:patient_plot}(a), four cubic regions of interests are indicated by squares with different colors. Each cube is 10 mm by 10 mm in the x-y direction and 14 mm in the z direction. The blue and red ROIs are in the adipose tissue, and the purple and orange ROIs are in muscle. This analysis is based on the assumption that the properties of the adipose and muscle of the object is close to expected. Since the elemental composition of adipose could vary greatly from person to person, we compared the attenuation of patient in adipose ROI to attenuation coefficients corresponding to the three representative adipose compositions with varying lipid content \cite{White}. 

In figure \ref{fig:patient_plot}(b) from left to right are the plots with the ground truth of adipose with different lipid percentages and muscle, respectively. The first three columns of plots are calculated using the same set of ROIs but different ground truths.
No matter which composition of adipose is used, the overall performance of MAR-DEAM on ROI1 is always better than the performance of the original DEAM. If adipose with 87.3\% lipid is used as the ground truth, the bias of MAR-DEAM is within +/-2\% for energies greater than 60 keV, and the MAE of MAR-DEAM is within 4\% after 60 keV. In this case, the percentage MAE of ROI1 is reduced by approximately a factor of 5 by the proposed method.
MAR-DEAM still performs better than the original DEAM on the regions of muscles. The percentage MAE of muscle region 1 is reduced by approximately a factor of 8 by MAR-DEAM for energies greater than 60 keV, and the percentage MAE of muscle region 2 is reduced by approximately a factor of 5 for energies greater than 80 keV.

\section{Conclusions}

In this work, we proposed a MAR framework for iterative DECT algorithms. DEAM was taken as an example of the iterative DECT algorithm. DEAM with the proposed MAR framework outperforms the original DEAM and the vendor reconstruction with O-MAR on the experimental test. The performance of the algorithm is also quantitatively analyzed on both the phantom and the patient data, with respect to estimating the attenuation coefficient at energies from 20 keV to 150 keV. The results show that the proposed MAR framework could significantly reduce the percentage bias and MAE in the selected regions.

\section{Acknowledgments} 
This study is supported by NIH R01 CA 212638 and Imaging Sciences Pathway T32EB014855(MM) from US National Institutes of Health. We thank the Alvin J. Siteman Cancer Center at Washington University School of Medicine for their help with clinical data acquisition.



\small
\bibliographystyle{unsrt}
\bibliography{ref}

\end{document}